\documentclass[a4paper]{article}
\usepackage{CJKutf8}
\usepackage{cite}
\usepackage{INTERSPEECH2022}
\usepackage[hyphens]{url}
\usepackage{hyperref}
\usepackage{makecell}
\usepackage{svg}
\usepackage{color}
\usepackage{multirow}
\usepackage{url}

\title{Automatic Prosody Annotation with Pre-Trained Text-Speech Model}
\name{Ziqian Dai$^{1,2}$, Jianwei Yu$^1$, Yan Wang$^1$, Nuo Chen$^{1,2}$, Yanyao Bian$^1$, Guangzhi Li$^1$, \\ Deng Cai, Dong Yu$^1$}
\address{
  $^1$Tencent AI Lab,
  $^2$Peking University 
  }
\email{daizq@stu.pku.edu.cn, tomasyu@tencent.com, brandenwang@tencent.com, dyu@tencent.com}

\begin{document}
\ninept
\maketitle
\begin{abstract}
  Prosodic boundary plays an important role in text-to-speech synthesis (TTS) in terms of naturalness and readability. However, the acquisition of prosodic boundary labels relies on manual annotation, which is costly and time-consuming. In this paper, we propose to automatically extract prosodic boundary labels from text-audio data via a neural text-speech model with pre-trained audio encoders. This model is pre-trained on text and speech data separately and jointly fine-tuned on TTS data in a triplet format: \{speech, text, prosody\}. The experimental results on both automatic evaluation and human evaluation demonstrate that: 1) the proposed text-speech prosody annotation framework significantly outperforms text-only baselines;  2) the quality of automatic prosodic boundary annotations is comparable to human annotations; 3) TTS systems trained with model-annotated boundaries are slightly better than systems that use manual ones. Code is released\footnote{\url{https://github.com/Daisyqk/Automatic-Prosody-Annotation}}.
\end{abstract}

\section{Introduction}
\begin{figure}[htb]     
  \center{\includegraphics[width=0.45\textwidth]{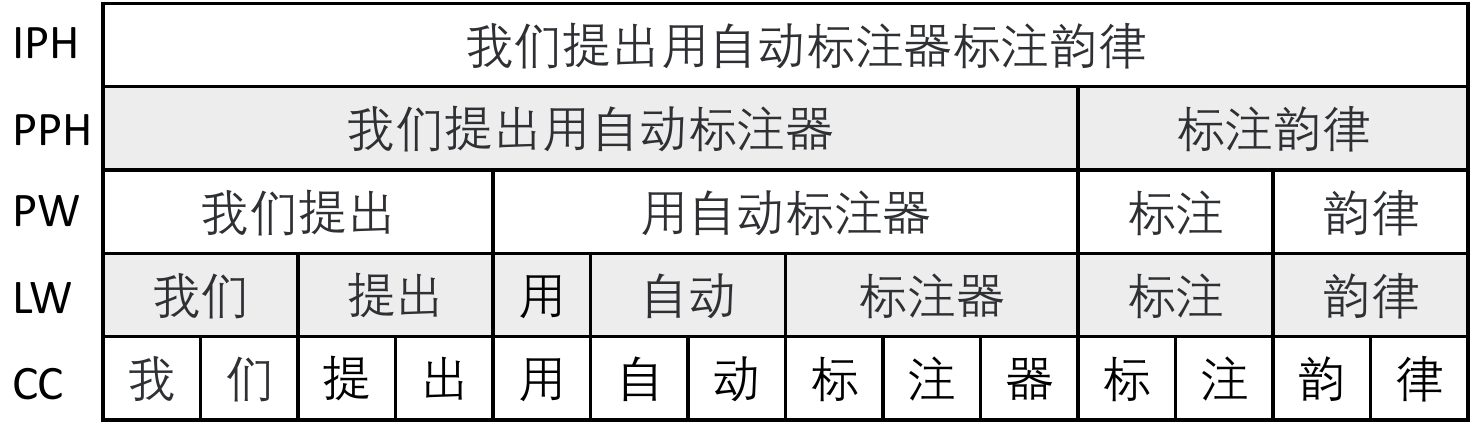}}        
  \caption{Five-level tree structure of prosodic boundary}
  \label{fig:Figure_1}
\end{figure}

% 为啥要韵律标椎
In text-to-speech synthesis (TTS), prosody modeling plays an important role in synthesizing high naturalness and intelligibility speech.
Due to the scarcity and high cost of prosody annotation in the current TTS dataset, there have been a lot of works attempting to model the prosody in a latent space without explicit prosody annotation \cite{wang2018style, liu2021expressive}.
However, recent research efforts have shown that using explicit hierarchical prosodic boundary annotation \cite{pan2019mandarin} in training and inference can still improve the fidelity and expressiveness of Mandarin speech synthesis \cite{yu2020durian}, which indicates that prosody annotation is still useful for TTS system construction.
As shown in Figure~\ref{fig:Figure_1}, the hierarchical prosody annotation adopted in this work categorizes the prosodic boundaries of Mandarin speech into five levels, including Character (CC), Lexicon Word (LW), Prosodic Word (PW), Prosodic Phrase (PPH) and Intonational Phrase (IPH)~\cite{2018Mandarin}.

\vspace{-0.1cm}
% Prosodic boundary is an import component which is required in both system training and inference, which greatly improves the naturalness and expressiveness of speech~\cite{2012Prosody}.

\begin{figure}[t]     
  \center{\includegraphics[width=0.4\textwidth] {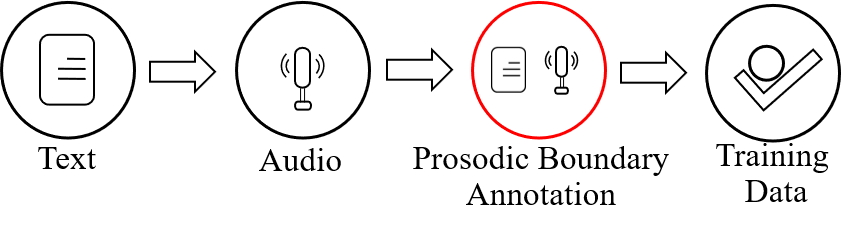}}
  \caption{ The pipeline of TTS training data collection. Prosodic boundary annotation is done by human annotators. }
  \vspace{-0.5cm}
  \label{fig:data_annotation}
\end{figure}

% 人工韵律标注的问题
While the importance of prosodic boundary has been demonstrated in previous studies, one most critical challenge is to obtain the prosodic boundaries.
The pipeline for collecting TTS training data with prosody annotations is shown in Figure~\ref{fig:data_annotation}, the acquisition of prosodic boundaries relies on manual annotation with text and speech, which is expensive and time-consuming. 
In addition, through preliminary experiments, we also find the inter-annotator agreement between different human annotators is low, indicating that prosody annotation can be ambiguous and the inconsistency may lead to difficulties in training models.

In this paper, we propose to reduce the cost of prosody annotation and improve the label consistency via an automatic annotator. Our key idea is to automatically annotate the prosodic boundaries through a pre-trained text-speech model that takes a text-speech pair as input. 
Specifically, the proposed model consists of three components: a text encoder, an audio encoder and a multi-modal fusion decoder. The former two components are pre-trained on text and audio data respectively, and the multi-modal fusion decoder is optimized using triplet format TTS data: \{speech, text, prosody\}. 

Three experiments are conducted to evaluate the effectiveness of this automatic annotator.
The first one directly calculates the precision, recall and f1 scores of the automatic annotations, the second one compares the accuracy of human- and model-annotated boundaries using A/B test, while the third evaluates the naturalness of TTS systems trained with different prosodic boundaries. 
Surprisingly, the experimental results show that the model-annotated boundaries do not result in worse, but instead slightly better, TTS performance. 
We attribute the results to our previous finding that human-annotated prosodic boundaries are inconsistent across individuals. 
% It partially verifies our previous findings that the human-annotated prosody labels are not perfect, they are inconsistent across individuals 

The contributions of this paper are listed as follows: 
1) We propose a text-speech framework to automatically annotate prosodic boundaries. 
2) Different kinds of audio encoders pre-trained on character-level targets and phonetic posteriorgram (PPG) \cite{sun2016phonetic} are systematically investigated to improve the annotation quality. 
3) The experimental result suggests that automatic annotations can translate to comparable TTS performance with manual annotations.

% \section{Related Work}

% \subsection{Text-based prosody estimation}
% A lot of studies have been carried out on predicting prosodic boundary based upon text. In early times, many traditional statistical methods have been explored together with manually designed features. Such as conditional random fields (CRF) \cite{2008Automatic,2010Automatic}, maximum entropy model (ME) \cite{DBLP:conf/interspeech/LiHW04,DBLP:conf/iscslp/LiuJT08,2011Polyphone}, hidden Markov model (HMM) \cite{2003Automatic} and classification and regression tree (CART) \cite{DBLP:conf/naacl/WangH91}. Later, deep recurrent neural network (RNN) was investigated due to its capability at capturing long-term context information \cite{DBLP:conf/interspeech/ZhengLWDT16,DBLP:conf/interspeech/HuangWLMC17,DBLP:conf/interspeech/ZhengTWL18}. And embedding features taking place of traditional linguistic features. Self-attention was also explored because it can connect two arbitrary characters directly regardless of distance\cite{2019Self}. More recently, with the development of natural language processing (NLP) pre-trained language models like BERT \cite{devlin2018bert}, these models are used to extract embedding features for various down-stream tasks including prosody prediction \cite{DBLP:conf/nodalida/TalmanSCKTV19,DBLP:journals/corr/abs-2012-15404,2019Pre}, which has brought a significant improvement in performance due to their strong ability to capture contextual semantic and syntactic information.

\section{Automatic Prosody Annotator}
\begin{figure}[htp]
    \centering
    \includegraphics[width=8cm]{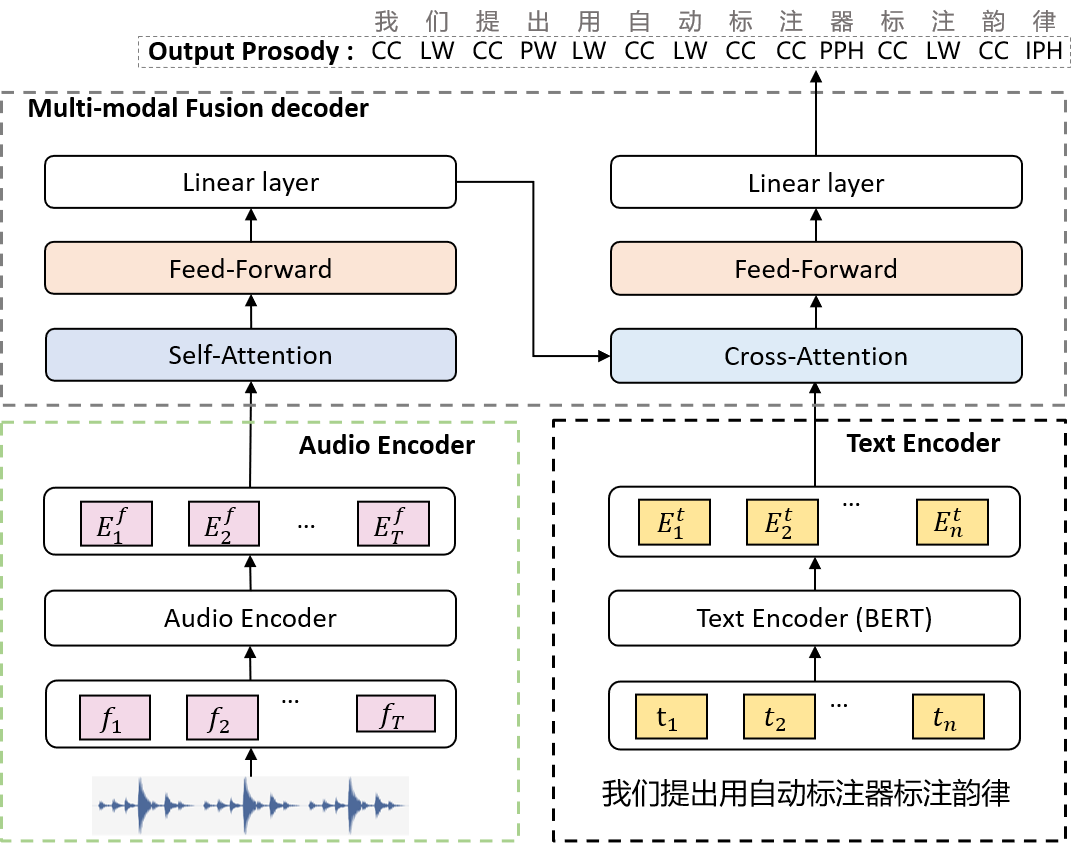}
    \caption{Architecture of the proposed text-speech automatic prosody annotation model.}
    \label{fig2:framework}
\end{figure}

In this section, we introduce our proposed automatic prosody annotator framework. Similar to human annotation, the model is requested to annotate prosodic boundaries according to the prosody information inherently contained in audio, so it takes both audio waveform and text as input. 
This is also the main difference between our model and those text-based prosodic boundary prediction models~\cite{du2019automatic,ning2019chinese,du2019prosodic,sloan2019prosody,yan2020mandarin,2019Self,devlin2018bert, DBLP:conf/nodalida/TalmanSCKTV19,2019Pre, zhang2020unified,bai2021universal,pan2019mandarin,wang2021predicting}. 
As shown in Fig~\ref{fig2:framework}, the proposed framework consists of three main components:  a text encoder, an audio encoder, and a multi-modal fusion decoder.
The text and audio encoders are used to extract high-level hidden representations from text and audio respectively, and the multi-modal fusion decoder is used to fuse the text and audio information to estimate prosodic boundary.
Since this model requires paired audio-prosody training data that are sparse and costly, we initialize our encoders with pre-trained models.
% Due to the limited amount of annotated audio-prosody data, pre-trained models are adopted as the text and audio encoders in this work.
% Due to the limited amount of annotated audio-prosody data and the great success of pre-trained model 

% We propose to adopt an automatic annotator as the substitution of human annotators. The automatic annotator is a sequence labeling model that takes paired text and speech as input and outputs a sequence of prosody labels. The main framework of our model is shown in Figure  \ref{fig2:framework}, it consists of three main components:  text encoder, audio encoder and a multi-modal decoder. To take full advantage of the capabilities of pre-trained models, we used two large pre-trained models to initialize the parameters of the text and audio encoder, respectively. The multi-modal decoder, on the other hand, is lightweight and we train it from scratch. Finally, the whole model is fine-tuned on triple-format data: \{speech, text, prosody\}.

\vspace{-0.5cm}
\subsection{Text Encoder}
% Benefited from pre-training on large amount of data to predict randomly masked words and next sentence,  BERT can generate universal text embeddings containing rich contextual semantic information for various down-stream tasks like sequence labeling\cite{tsai2019small}. 
We apply a pre-trained Chinese Bert~\cite{devlin2018bert} as text encoder to embed each token, which refers to the character unit in Chinese, into a fixed-length vector containing contextual information.  

\subsection{Audio Encoder}
As prosody information is inherently contained in the audio modality, a powerful audio encoder that can extract representative prosody-related information from the audio waveform is important to boost estimation accuracy.

\noindent{\textbf{PPG extractor}}: Motivated by the great success of Phonetic posteriorgram (PPG) \cite{sun2016phonetic} in voice conversion, we firstly adopt a pre-trained conformer-based PPG extractor as the audio encoder. 
Specifically, the PPG extractor used in this work is a speaker-independent frame-level classifier that maps each input time frame to posterior probabilities of phonetic classes.
In automatic prosody annotator task, we believe that PPG can represent the duration and transition information of each phoneme in the audio, which is important to prosodic boundary estimation.
% More than 10k hour of multi-speaker ASR data are used to train the PPG extractor, which leads the model to be speaker independent.

\noindent{\textbf{Character-based encoder}}: However, as prosodic boundaries are usually related to the word and character language information, the phoneme-based PPG model misses the character-level language context information, which can lead to sub-optimal performance.
For example, these two sequences {\begin{CJK}{UTF8}{gbsn} "大学生物，必修课" and "大学生, 务必修课" have the same phone sequence, however, their prosodic boundaries are different. \end{CJK}} 
To this end, two character-based ASR models based on CNN and conformer \cite{gulati2020conformer} architectures are investigated.
Different from the PPG extractor, the character-based model transforms each input time frame to posterior probabilities of character classes, which can better keep the character-level information.
The difference between the two model architectures is that the CNN-based model focuses on the local information, while the conformer-based model considers the whole utterance information.
Note that, since there are more than 5000 characters in the vocabulary, the 512 dimensional hidden representations of the second last layer are adopted as the audio encoder hidden outputs for character-based pre-trained audio encoders.

% The prosody information is inherently contained in the audio modality. It is natural to think about extracting prosody related features from raw audio for further prediction of prosodic boundary. We try following three kinds of encoders for audio. \wy{All of them are pre-trained on large amount of automatic speech recognition(ASR) data.}

% \textbf{CNN Char} 
%  We firstly try CNN Char ASR model to embed raw audio. It is small scale and fast, but has no sequence modeling unit, thus can not extract contextual information. 

% \textbf{Conformer-Char}
%  Considering the influence of context information on prosody, we employ Conformer-Char ASR model to extract features containing contextual information.

% \textbf{Conformer-ppg}
% A Phonetic PosteriorGrams(PPG) is a time-versus-class matrix representing the posterior probabilities of each phonetic class for each specific time frame of one utterance\cite{2011Polyphone}. We use pre-trained model to extract PPG to align each frame of speech to text. Unlike the above two models which modeling on the unit of char, PPG is modeled on the unit of phone. 

\subsection{Multi-Modal Fusion Decoder}
With the audio and text hidden representations, a multi-modal fusion decoder is needed to fuse these two representations and then estimate the prosodic boundary.
One challenge of the fusion of these two hidden representations is that the frame-level audio hidden representation is usually much longer than the token-level text representation.
To address this issue, a cross-attention based multi-modal fusion decoder is proposed in this work.

As shown in Fig.3, the audio hidden representations are fed into a stack of 6 identical layers, each layer is composed of a multi-head self-attention layer followed by a feed-forward layer. Then, the transformed representations go through a linear layer, which changes their dimension to be the same with the text hidden representations. Lastly, the hidden representations of the text and audio modalities are fed into another stack of 6 identical layers, each composing a multi-head cross-attention layer followed by a feed-forward layer.

%the text hidden representations are forward into self-attention module, in the mean time, the audio hidden representations are forward into a {\color{red}{transformer block???}} followed by a linear layer.

Since the length of output prosodic boundary estimation is the same as the length of the input text sequence, in cross-attention, the text modality is used as query, the audio modality is used as key and value.
Let ${\bf{O}} = [{\bf{o_1}}, ..., {\bf{o_T}}] \in \bf\rm{R}^{T\times D}$ and ${\bf{X}} = [{\bf{x_1}}, ..., {\bf{x_N}}]\in \bf\rm{R}^{N\times D}$ denote the audio and text hidden input of the multi-head cross-attention layer respectively, where $D$ is the dimension of the hidden representation. The fusion of the audio and text modality can be expressed as follows:
\begin{align}
    {\bf{Q}}_{X}, {\bf{K}}_{O}, {\bf{V}}_{O} &= {\bf{W}}_{Q}{\bf{X}},  {\bf{W}}_{K}{\bf{O}}, {\bf{W}}_{V}{\bf{O}} \\
     {\bf{H}}&= {\text{softmax}}(\frac{{\bf{Q}}_{X}{\bf{K}}_{O}^T}{\sqrt{D}}){\bf{V}}_{O}
\end{align}
where ${\bf{H}} \in  \bf\rm{R}^{N\times D}$ is the fused hidden output of the cross-attention layer for prosodic boundary estimation, and $\bf{W}_{Q}$, $\bf{W}_{K}$ and $\bf{W}_{V}$ are trainable matrices, separately.
Such a cross attention layer is designed to allow the model to automatically learn the alignment of the audio and text hidden inputs. 

\subsection{Training Objective}
The fused hidden vector ${\bf{H}}$ is then fed into an output linear layer with \textit{softmax} function to obtain the probability distribution of prosodic boundaries. 
Cross Entropy (CE) criterion is adopted as the training objective of the proposed model:
\begin{align}
    { \boldsymbol{p}_k = {\text{softmax}}(\textbf{WH} + \textbf{b})} \\
    {\mathcal{L}_{\text{ce}} = \sum_{k=1}^{N} -y_k \log \boldsymbol{p}_k }
\end{align}
where $\boldsymbol{p}_k \in \bf\rm{R}^{1\times |L|}$ represents the probability of the $k$-th token in the input sequence, $L$ is the annotation labeling set, $\textbf{W} \in \bf\rm{R}^{|L| \times D}$ and $\textbf{b}$ are the model weights and bias of the output layer.  
\begin{table*}[htb]
%\caption{Results of text-based and different audio-based models. "Pre-train" indicates using pre-trained audio encoders with fixed model parameters and "Fine-tune" indicates further fine-tune the entire model using the prosody annotation data. "pre.", "rec." and "f1" denotes the precision, recall, and f1 score respectively.}
%\caption{Results of text-based and different audio-based models. Note that fine-tuned Bert is used in all the models in this table.  Model \#3 and \#6 are trained from scratch with random initialized audio-encoder, Model \#4, \#7, \#9 are trained using pre-trained audio encoders with fixed model parameters, Model \#5 and \#8 have pre-trained audio encoders further fine-tuned on the prosody annotation data. }
%\caption{Results of text-based and different audio-based models. In column "Pre-train", $\times$ indicates audio encoder is trained from scratch with random initialization, $\surd$ indicates pre-trained but fixed during fine-tuning. $\surd$ in column "Fine-tune" indicates the pre-trained audio encoder is further fine-tuned.}
\caption{Results of text-based and different audio-based models. Bert is pre-trained and fine-tuned in all the models. Column "Pre-trained" indicates whether the audio-encoder is pre-trained. Column "Fixed" indicates whether the audio-encoder is fixed during training. "pre.", "rec." and "f1" denotes the precision, recall, and f1 score respectively.}
    \centering
    \scalebox{0.85}{
    \begin{tabular}{c|c|cc|ccc|ccc|ccc|ccccccccccc}
    \toprule
    \multirow{2}{*}{ID} &\multirow{2}{*}{Model}  & \multicolumn{2}{c|}{Audio encoder} &\multicolumn{3}{c|}{LW} & \multicolumn{3}{c|}{PW} &\multicolumn{3}{c|}{PPH} & \multicolumn{3}{c}{IPH} \\
     \cline{3-16}
     &     & \multirow{1}{*}{Pre-trained} & \multirow{1}{*}{Fixed} &pre. & rec. & f1 &pre. & rec. & f1&pre. & rec. & f1&pre. & rec. & f1 \\
     \hline
    %  1&\multirow{2}{*}{Bert}  & $\surd$ & $\times$&0.87 &0.90 & 0.89 &0.22& 0.16 &0.19& 0.77 &0.78 &0.78&1.00&0.99&1.00\\
     1&\multirow{1}{*}{Bert} &\multicolumn{2}{c|}{-} &0.87 &0.90 &0.89 &0.24 &0.21 &0.22 &0.80 &0.78 &0.79 &1.00 &0.99 &1.00\\
    %  2&                  & $\surd$ & $\surd$ &0.87 &0.90 &0.89 &0.24 &0.21 &0.22 &0.80 &0.78 &0.79 &1.00 &0.99 &1.00\\
     \hline
     \hline
     2&\multirow{1}{*}{Human}  &\multicolumn{2}{c|}{-}&0.91&0.92&0.91 &0.49& 0.45&0.45&0.83 &0.71&0.76&1.00&1.00&0.99\\

     \hline
     \hline
     3&\multirow{3}{*}{CNN-Char}  & $\times$  & $\times$ &0.89 &0.84 &0.86 &0.24 &0.48 &0.32 &0.89 &0.80 &0.84 &1.00 &0.98 &0.99\\
     4&&  $\surd$  & $\surd$ &0.88 &0.90 &0.89 &0.31 &0.35 &0.33 &0.88 &0.80 &0.84 &1.00 &0.99 &0.99\\
     5&&  $\surd$  & $\times$ &0.89 &0.91 &0.90 &0.39 &0.44 &0.41 &0.92 &0.83 &0.87 &1.00 &0.99 &0.99\\
     \hline
     6&\multirow{3}{*}{Conformer-Char}  & $\times$  & $\times$ &0.85 &0.90 &0.87 &0.20 &0.27 &0.23 &0.90 &0.64 &0.75 &0.99 &0.97 &0.99\\
     7&&  $\surd$  & $\surd$  &0.91 &0.91 &0.91 &0.47 &\bf0.65 &0.54 &\bf0.96 &\bf0.90 &\bf0.93 &1.00 &0.98 &0.99\\
     8&&  $\surd$  & $\times$ &0.9 &0.91 &0.91 &0.43 &0.56 &0.48 &0.94 &0.89 &0.91 &1.00 &1.00 &0.99\\
     \hline
     9&\multirow{1}{*}{Conformer-PPG}  & $\surd$  & $\surd$ &\bf0.92&\bf0.92&\bf0.92&\textbf{0.51}&0.64&\textbf{0.57}&0.95&0.89&\textbf{0.92}&1.00&0.99&1.00\\
    %   & $\times$  & $\surd$  & $\surd$ &\bf0.92 &\bf0.92 &\bf0.92 &\bf0.54 &0.63 &\bf0.58 &0.94 &\bf0.90 &0.92 &1.00 &1.00 &0.99\\
    
    \bottomrule
    \end{tabular}
    }
    \label{table_1}
\end{table*}
\vspace{-0.5cm}
\section{Experimental Setup}
The dataset, implementation details and the baselines of this work are introduced in this section. 
% \red{We} evaluate the proposed automatic annotator on both automatic metrics and human evaluation. In Section~\ref{sec: automatic_evaluation}, we report the recall and precision of different methods. The performance of the model was then compared in depth with human performance, and two types of human evaluations were further performed: the first one is an A/B test that directly compares the quality of human and automatic boundaries. 
% In the second evaluation, the Mean Opinion Score (MOS) test between TTS systems trained on automatic and manual prosodic boundaries is given.
% The experiment in this work is organized as follows:
% First, we compare the prosodic boundary annotation performance of the proposed methods using different types of audio encoder and optimization approach.
% Then, to show the quality of model-annotated labels, a comparison between human and model annotation based on human evaluation is given.
% Finally, to verify that the automatic annotation generated by proposed model is applicable to TTS system construction, the Mean Opinion Score (MOS) test between TTS systems trained on automatic and manual prosody labels is given. 
% This section briefly introduces the organization of the dataset and the implementation details.

\subsection{Dataset}
\noindent{\bf{Train and dev sets}}: In this work, around 12.2k utterances ($\approx$160 hour) recorded by 28 different speakers are used to construct the proposed automatic prosody annotation models, where 95\% of the data is used as the training set and the rest 5\% is used as the dev set. 
Note that human prosody annotation is used as the training target. \\
\noindent{\bf{Evaluation set}}:
To evaluate the performance of the proposed models, 5.9k utterances ($\approx$8.8 hour) from another 9 speakers are used as the test set. 
Note that these speakers are unseen in training.

\subsection{Implementation details}
% \subsubsection{Text Encoder}
\noindent{\bf{Text Encoder}}:
% The Chinese BERT we use is pre-trained by ourselves internally. It consists of 12 transformer layers, the dimension of hidden size is 768, and the number of attention heads is 12. It can be replaced by bert-base-chinese model on Hugging Face, but the F1 score of prediction will drop slightly with 0.01 on average. \\
We use an internal Chinese \textit{BERT-base} pre-trained on a 300 GB news corpus as our text encoder.  

% \subsubsection{Audio encoder}
\noindent{\bf{Pre-trained Audio encoder}}: All the audio encoders used in this work are pre-trained using open sourced 10k hour WenetSpeech \cite{zhang2021wenetspeech} dataset. 80 dimension FBank and 3 dimension pitch are concatenated as the input feature. All these models are optimized using Adam optimizer \cite{2014Adam} for 50 epochs. \\
% \noindent{\bf{PPG-base Encoder}}: 
\noindent{1) PPG-based Encoder}: 
A conformer-based model consisting of 2 convolutional layers and 12 conformer blocks are adopted to build the pre-trained PPG-based encoder. Details can be found in \cite{tian2022improving} \footnote{\url{https://github.com/jctian98/e2e_lfmmi/blob/master/egs/aishell1/conf/tuning/train_pytorch_conformer_kernel15.yaml}}.
The frame-level context independent phone (218 phones) alignment generated from a GMM-HMM model~ \footnote{\url{https://github.com/wenet-e2e/WenetSpeech/blob/main/toolkits/kaldi/run.sh}} is used as the training target. 
The model is optimized using a frame-level cross-entropy objective. \\
% \noindent{\bf{Character-base Encoder}}:
\noindent{2) Character-based Encoder}:
The CNN Character-based encoder only contains two 2D convolution layers and a linear output layer. The conformer-based encoder uses the same model architecture as the PPG encoder, except the final output layer.
The character based audio encoder is optimized using CTC cost function in an End-to-End manner.
The number of the Chinese Character is 5546.\\
% \subsubsection{Multi-modal fusion layer}
\noindent{\bf{Multi-modal fusion decoder}}:
The self-attention and cross-attention layers we use both have 8 heads. The feed-forward layer composes of two linear layers with a ReLU activation\cite{2010Rectified}. A residual connection\cite{he2015deep} is employed around each self-attention, cross-attention and feed forward layer. 
% Layer normalization\cite{ba2016layer} is applied to stabilize the activation of the network after residual connection. What's more, to avoid over-fitting, dropout\cite{2014Dropout} layers with the keep probability of 0.9 are added before the residual connections, after softmax in self-attention layers and cross-attention layers, and after RELU in feed-forward layers. 
Parameter optimization is performed using
Adam optimizer with learning rate of 0.0001. \\
% \subsubsection{DurAIN TTS}
\noindent{\bf{DurAIN TTS}}:
In this work, the DurIAN \cite{yu2020durian} TTS framework is used to verify the quality of the estimated prosodic boundary for TTS system construction. 
Specifically, 90\% of the evaluation set is used to train the DurIAN TTS acoustic model and HiFi-GAN vocoder \cite{kong2020hifi} and the other 10\% of this set is used for evaluation. 
Both the acoustic model and vocoder are optimized using Adam optimizer for 100k iterations with batch size of 32  using 1 V100 GPU.  
Details of the DurIAN TTS framework can be found in \cite{yu2020durian}.

\subsection{Baselines}
There are two baselines in our experiment, the first one is \textbf{BERT}, which shares the same architecture with our text encoder, and only takes text as input. The second one is \textbf{human performance}, in which we recruite seven annotators to annotate the test set and report their average performance as a proxy for human performance. 

% yu2020durian

\section{Experimental Results}
We evaluate the proposed automatic annotator on both automatic metrics and human evaluation. In Section~\ref{sec: automatic_evaluation}, we report the recall and precision of different methods. The model performance is then compared in depth with human performance, and two types of human evaluations are performed: the first one is an A/B test that directly compares the quality of human and automatic boundaries. 
In the second evaluation, the Mean Opinion Score (MOS) test between TTS systems trained on automatic and manual prosodic boundaries is given.
% The experimental results conducted on automatic and humman evaluation are presented in this section. 
% \subsection{Automatic prosody estimation}
\subsection{Automatic Evaluation}
\label{sec: automatic_evaluation}
\begin{figure}[htb]
    \centering
    \includegraphics[width=8cm]{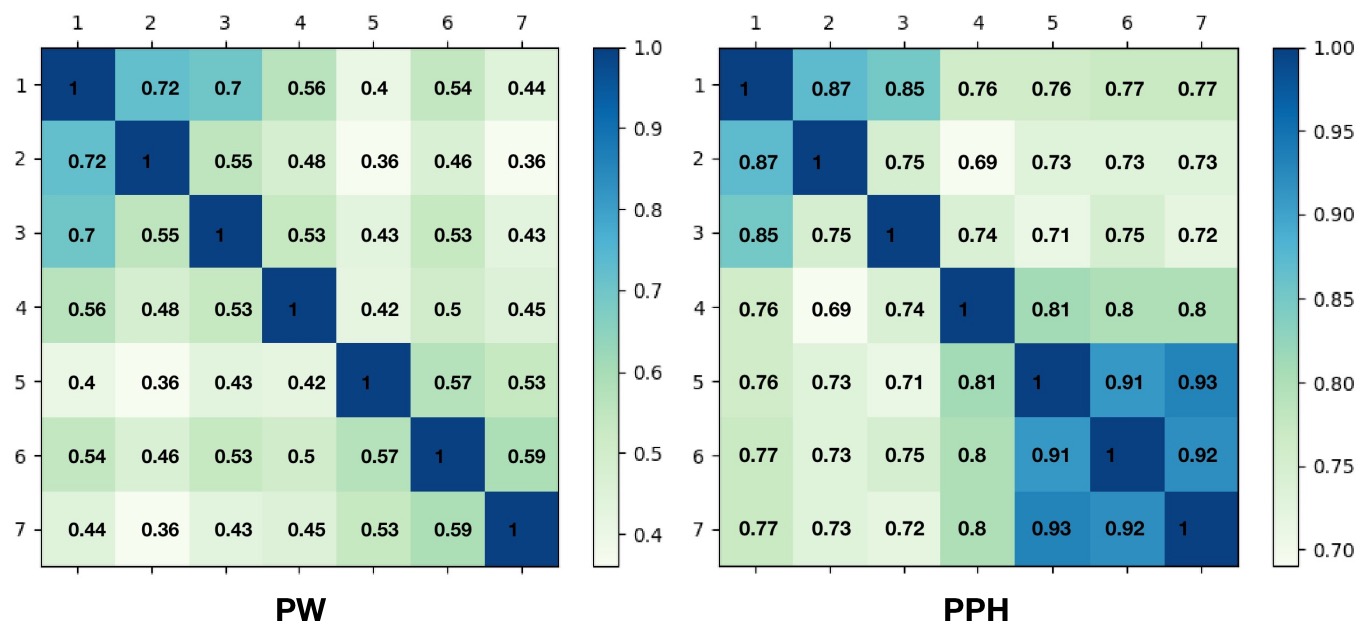}
    \caption{Kappa coefficient between 7 human annotators on PW and PPH prosodic boundary annotation.}
    \label{fig4:kappa}
\end{figure}

We report the recall, precision, and f1 score of different methods on different boundary levels in Table~\ref{table_1}. We can see that most models have an f1 score of 0.9 or higher for both LW and IPH, meaning they are not difficult for humans and models. Therefore, in our experiments, we focus on two other boundaries, PW and PPH. We observe several interesting trends:\\
\noindent 1) All multi-modal methods perform much better than the text-only model, BERT (Model \#1 v.s. Model \#3 - \#9). It confirms that the proposed model can effectively utilize audio modality to discriminate the prosodic boundaries of PW and PPH. \\
\noindent2) Pre-training matters on larger models. CNN-Char performs better than Conformer-Char without pre-training (Model \#3 v.s. Model \#6), while the opposite was true after pre-training (Model\#4-5 v.s. Model \#7-8), indicating that the conformer model with much more parameters requires a large amount of pre-training data to optimize. Furthermore, once the model size is large enough, there is no need for fine-tuning, as it can lead to over-fitting (Model \#7 v.s. Model \#8).\\
\noindent 3) Models with conformer-based audio encoders significantly outperform models using CNN-based audio encoder (Model \#5 - \#6 vs Model \#7 - \#9). This can be attributed to not only the larger model size but also the modeling of long context prosody information. In addition, the PPG-based conformer audio-encoder also shows comparable results with the character-based conformer encoder (Model \#7 v.s. Model \#9).   \\
\noindent 4) Surprisingly, the proposed model shows higher precision, recall, and f1 score over the human annotators (Model \#2 v.s. Model \#7, 9). However, this only means that models are more consistent with the prosody annotation given in the original dataset than the human annotators hired in this experiment. This also indicates that the sense of prosodic boundary, especially PW and PPH, can be different across individuals. 
To verify this assumption, we calculate the Fleiss’ Kappa coefficients~\cite{landis1977measurement} among human annotators. As Figure~\ref{fig4:kappa} shows, the Kappa coefficients between most annotators are less than 0.6 on PW prosodic boundary, which means they are only ``moderately consistent''. For IPH, the Kappa coefficients can be relatively higher, however, many of them are still below 0.8.
In addition, the annotators also report that the difference between PW and IPH can be ambiguous during the annotation process.

\vspace{-0.5cm}

\subsection{Human Evaluation}
\label{sec:human_evaluation}
In section~\ref{sec: automatic_evaluation}, we find the inconsistency problem exists in human annotations. it raises, to some extent, a sense of mistrust to the result of automatic evaluation. Therefore, we further conduct two human evaluations to fairly evaluate the performance of the proposed automatic annotator. 

\subsubsection{A/B test}
To compare the quality between human and automatic prosody annotation, we first randomly sample 300 utterances from the test set whose automatic prosody annotations generated by the PPG model (Model \#9) are different from the prosody annotation provided in the dataset. We hire 3 professional audio annotators from a third-party company to compare these two annotations. These annotators are proficient in audio tasks but know nothing about the models. For each utterance, they are asked to refer to the given audio record and select the better one from the manual and automatic annotation in an A/B test manner. The order of the annotations for each utterance is randomly shuffled to exclude the bias. 

The results of this test show that: on \textbf{51\% (153)} utterances,  the automatic annotation gets more votes than the original human annotation in the dataset, which indicates that the proposed automatic prosody annotation method can be comparable to human annotators.
\begin{table}[h]
\begin{center}
\caption{MOS test results for TTS systems trained using different prosody annotations with 95\% confidence intervals.}
 \scalebox{0.85}{
\begin{tabular}{l|c}
\toprule
 Prosody Annotation & MOS\\
\hline
Automatic &3.890 $\pm$ 0.037\\
Manual  & 3.808 $\pm$ 0.032\\
NA &2.799 $\pm$ 0.035\\
\bottomrule
\end{tabular}}
\vspace{-1cm}
\end{center}
\end{table}

\subsubsection{TTS MOS Test}
The primary motivation of this work is to reduce the annotation cost of TTS system construction. Therefore, whether the automatic annotation is sufficient as an alternative to human annotation in TTS system training is worth studying. In this section, we take the DurAIN TTS as our test-bed and conduct crowd-sourced MOS tests to compare TTS systems trained with automatic prosody annotations, with manual annotations, and without prosody annotations~\footnote{Audio samples can be found in \url{https://daisyqk.github.io/Automatic-Prosody-Annotation_w/}}.  

For all TTS systems, we adopt the same text and prosody content in the original test set as inputs and randomly shuffle the order of the utterances to exclude other interference factors but only examine the audio prosody.  Note that each input used in the MOS test contains at least one PW or IPH prosodic boundary with at least 12 Chinese characters. For the system trained without prosodic boundary, the prosodic boundary in the input text will be omitted. Each audio sample is rated by 24 testers, who are asked to evaluate the prosody naturalness of the synthesized speech on a five-point scale, with the lowest and highest scores being 1 (“Bad”) and 5 (“Excellent”).

\noindent The MOS result in table 2 reveals that: \\
 1) Any type of prosodic boundary can significantly improve the naturalness of the TTS system.\\
 2) The TTS system trained using automatic boundaries slightly outperforms the system using manual ones. This result is in line with the result of A/B Test. 
As discussed in section~\ref{sec: automatic_evaluation}, since human annotators can be inconsistent and the prosody of the TTS data is usually annotated by multiple annotators, the inconsistency of different annotators will confuse the TTS model to learn to model the prosody information during training. In contrast, the automatic annotators are consistent over samples and thus leading to a more well-trained and natural TTS system.

\section{Conclusion}
In this paper, we propose a speech-text model to automatically annotate prosodic boundaries.
We separately pre-train text and audio encoders on large amounts of text and TTS data, then fine-tune them together with a multi-modal fusion decoder on TTS dataset. In experiments, beyond the automatic evaluation, we also conduct human evaluations to understand the proposed method more comprehensively.
% , which are manually annotating prosody and compare with automatic prediction, voting on manual and automatic labels according to the human recorded speech, training TTS model with different prosody and compare the naturalness of generated speech. 
Experimental results demonstrate that the proposed method shows comparable performance against human annotators, which shows the potential to use automatic prosody annotators to replace human annotators.

\bibliographystyle{IEEEtran}
\bibliography{mybib}

% Generated by IEEEtran.bst, version: 1.14 (2015/08/26)
\begin{thebibliography}{10}
\providecommand{\url}[1]{#1}
\csname url@samestyle\endcsname
\providecommand{\newblock}{\relax}
\providecommand{\bibinfo}[2]{#2}
\providecommand{\BIBentrySTDinterwordspacing}{\spaceskip=0pt\relax}
\providecommand{\BIBentryALTinterwordstretchfactor}{4}
\providecommand{\BIBentryALTinterwordspacing}{\spaceskip=\fontdimen2\font plus
\BIBentryALTinterwordstretchfactor\fontdimen3\font minus
  \fontdimen4\font\relax}
\providecommand{\BIBforeignlanguage}[2]{{%
\expandafter\ifx\csname l@#1\endcsname\relax
\typeout{** WARNING: IEEEtran.bst: No hyphenation pattern has been}%
\typeout{** loaded for the language `#1'. Using the pattern for}%
\typeout{** the default language instead.}%
\else
\language=\csname l@#1\endcsname
\fi
#2}}
\providecommand{\BIBdecl}{\relax}
\BIBdecl

\bibitem{wang2018style}
Y.~Wang, D.~Stanton, Y.~Zhang, R.-S. Ryan, E.~Battenberg, J.~Shor, Y.~Xiao,
  Y.~Jia, F.~Ren, and R.~A. Saurous, ``Style tokens: Unsupervised style
  modeling, control and transfer in end-to-end speech synthesis,'' in
  \emph{International Conference on Machine Learning}.\hskip 1em plus 0.5em
  minus 0.4em\relax PMLR, 2018, pp. 5180--5189.

\bibitem{liu2021expressive}
R.~Liu, B.~Sisman, G.~Gao, and H.~Li, ``Expressive tts training with frame and
  style reconstruction loss,'' \emph{IEEE/ACM Transactions on Audio, Speech,
  and Language Processing}, vol.~29, pp. 1806--1818, 2021.

\bibitem{pan2019mandarin}
H.~Pan, X.~Li, and Z.~Huang, ``A mandarin prosodic boundary prediction model
  based on multi-task learning.'' in \emph{Interspeech}, 2019, pp. 4485--4488.

\bibitem{yu2020durian}
C.~Yu, H.~Lu, N.~Hu, M.~Yu, C.~Weng, K.~Xu, P.~Liu, D.~Tuo, S.~Kang, G.~Lei
  \emph{et~al.}, ``Durian: Duration informed attention network for speech
  synthesis,'' \emph{Proc. Interspeech 2020}, pp. 2027--2031, 2020.

\bibitem{2018Mandarin}
K.~Xie and P.~Wei, \emph{Mandarin Prosody Prediction Based on Attention
  Mechanism and Multi-model Ensemble}.\hskip 1em plus 0.5em minus 0.4em\relax
  Intelligent Computing Theories and Application, 2018.

\bibitem{sun2016phonetic}
L.~Sun, K.~Li, H.~Wang, S.~Kang, and H.~Meng, ``Phonetic posteriorgrams for
  many-to-one voice conversion without parallel data training,'' in \emph{2016
  IEEE International Conference on Multimedia and Expo (ICME)}.\hskip 1em plus
  0.5em minus 0.4em\relax IEEE, 2016, pp. 1--6.

\bibitem{du2019automatic}
Y.~Du, Z.~Wu, S.~Kang, D.~Su, D.~Yu, and H.~Meng, ``Automatic prosodic
  structure labeling using dnn-bgru-crf hybrid neural network,'' in \emph{2019
  Asia-Pacific Signal and Information Processing Association Annual Summit and
  Conference (APSIPA ASC)}.\hskip 1em plus 0.5em minus 0.4em\relax IEEE, 2019,
  pp. 1234--1238.

\bibitem{ning2019chinese}
M.~Ning, ``Chinese prosodic phrase prediction based on shallow semantic
  features,'' in \emph{2019 IEEE 5th International Conference on Computer and
  Communications (ICCC)}.\hskip 1em plus 0.5em minus 0.4em\relax IEEE, 2019,
  pp. 245--250.

\bibitem{du2019prosodic}
Y.~Du, Z.~Wu, S.~Kang, D.~Su, D.~Yu, and H.~Meng, ``Prosodic structure
  prediction using deep self-attention neural network,'' in \emph{2019
  Asia-Pacific Signal and Information Processing Association Annual Summit and
  Conference (APSIPA ASC)}.\hskip 1em plus 0.5em minus 0.4em\relax IEEE, 2019,
  pp. 320--324.

\bibitem{sloan2019prosody}
R.~Sloan, S.~S. Akhtar, B.~Li, R.~Shrivastava, A.~Gravano, and J.~Hirschberg,
  ``Prosody prediction from syntactic, lexical, and word embedding features,''
  in \emph{10th ISCA Speech Synthesis Workshop}, 2019.

\bibitem{yan2020mandarin}
Y.~Yan, J.~Jiang, and H.~Yang, ``Mandarin prosody boundary prediction based on
  sequence-to-sequence model,'' in \emph{2020 IEEE 4th Information Technology,
  Networking, Electronic and Automation Control Conference (ITNEC)},
  vol.~1.\hskip 1em plus 0.5em minus 0.4em\relax IEEE, 2020, pp. 1013--1017.

\bibitem{2019Self}
C.~Lu, P.~Zhang, and Y.~Yan, ``Self-attention based prosodic boundary
  prediction for chinese speech synthesis,'' in \emph{ICASSP 2019 - 2019 IEEE
  International Conference on Acoustics, Speech and Signal Processing
  (ICASSP)}, 2019.

\bibitem{devlin2018bert}
J.~Devlin, M.-W. Chang, K.~Lee, and K.~Toutanova, ``Bert: Pre-training of deep
  bidirectional transformers for language understanding,'' \emph{arXiv preprint
  arXiv:1810.04805}, 2018.

\bibitem{DBLP:conf/nodalida/TalmanSCKTV19}
\BIBentryALTinterwordspacing
A.~Talman, A.~Suni, H.~{\c{C}}elikkanat, S.~Kakouros, J.~Tiedemann, and
  M.~Vainio, ``Predicting prosodic prominence from text with pre-trained
  contextualized word representations,'' in \emph{Proceedings of the 22nd
  Nordic Conference on Computational Linguistics, NoDaLiDa 2019, Turku,
  Finland, September 30 - October 2, 2019}, M.~Hartmann and B.~Plank,
  Eds.\hskip 1em plus 0.5em minus 0.4em\relax Link{\"{o}}ping University
  Electronic Press, 2019, pp. 281--290. [Online]. Available:
  \url{https://aclanthology.org/W19-6129/}
\BIBentrySTDinterwordspacing

\bibitem{2019Pre}
B.~Yang, J.~Zhong, and S.~Liu, ``Pre-trained text representations for improving
  front-end text processing in mandarin text-to-speech synthesis,'' in
  \emph{Interspeech 2019}, 2019.

\bibitem{zhang2020unified}
Y.~Zhang, L.~Deng, and Y.~Wang, ``Unified mandarin tts front-end based on
  distilled bert model,'' \emph{arXiv preprint arXiv:2012.15404}, 2020.

\bibitem{bai2021universal}
Z.~Bai and B.~Hu, ``A universal bert-based front-end model for mandarin
  text-to-speech synthesis,'' in \emph{ICASSP 2021-2021 IEEE International
  Conference on Acoustics, Speech and Signal Processing (ICASSP)}.\hskip 1em
  plus 0.5em minus 0.4em\relax IEEE, 2021, pp. 6074--6078.

\bibitem{wang2021predicting}
Q.~Wang, X.~Wang, W.~Liu, and G.~Chen, ``Predicting the chinese poetry prosodic
  based on a developed bert model,'' in \emph{2021 IEEE 2nd International
  Conference on Big Data, Artificial Intelligence and Internet of Things
  Engineering (ICBAIE)}.\hskip 1em plus 0.5em minus 0.4em\relax IEEE, 2021, pp.
  583--586.

\bibitem{gulati2020conformer}
A.~Gulati, J.~Qin, C.-C. Chiu, N.~Parmar, Y.~Zhang, J.~Yu, W.~Han, S.~Wang,
  Z.~Zhang, Y.~Wu \emph{et~al.}, ``Conformer: Convolution-augmented transformer
  for speech recognition,'' \emph{arXiv preprint arXiv:2005.08100}, 2020.

\bibitem{zhang2021wenetspeech}
B.~Zhang, H.~Lv, P.~Guo, Q.~Shao, C.~Yang, L.~Xie, X.~Xu, H.~Bu, X.~Chen,
  C.~Zeng \emph{et~al.}, ``Wenetspeech: A 10000+ hours multi-domain mandarin
  corpus for speech recognition,'' \emph{arXiv preprint arXiv:2110.03370},
  2021.

\bibitem{2014Adam}
D.~P. Kingma and J.~Ba, ``Adam: A method for stochastic optimization,''
  \emph{arXiv e-prints}, 2014.

\bibitem{tian2022improving}
J.~Tian, J.~Yu, C.~Weng, Y.~Zou, and D.~Yu, ``Improving mandarin end-to-end
  speech recognition with word n-gram language model,'' \emph{arXiv preprint
  arXiv:2201.01995}, 2022.

\bibitem{2010Rectified}
V.~Nair and G.~E. Hinton, ``Rectified linear units improve restricted boltzmann
  machines vinod nair,'' in \emph{International Conference on International
  Conference on Machine Learning}, 2010.

\bibitem{he2015deep}
K.~He, X.~Zhang, S.~Ren, and J.~Sun, ``Deep residual learning for image
  recognition,'' 2015.

\bibitem{kong2020hifi}
J.~Kong, J.~Kim, and J.~Bae, ``Hifi-gan: Generative adversarial networks for
  efficient and high fidelity speech synthesis,'' \emph{Advances in Neural
  Information Processing Systems}, vol.~33, pp. 17\,022--17\,033, 2020.

\bibitem{landis1977measurement}
J.~R. Landis and G.~G. Koch, ``The measurement of observer agreement for
  categorical data,'' \emph{biometrics}, pp. 159--174, 1977.

\end{thebibliography}
\end{document}